"Sufficiently Advanced Technology" for Gravitational Wave Detection

for the Proceedings of the 2018 Les Houches Summer School on Gravitational Waves

by Peter R Saulson    (DCC number: LIGO-P1900203-v3)

ABSTRACT:

The science fiction writer Arthur C. Clarke wrote, "Any sufficiently advanced technology is indistinguishable from magic."[1] While not magical by any means, the technology used to detect gravitational waves starting in 2015 is surely sufficiently advanced to be remarkable by any ordinary standard. That technology was developed over a period of almost six decades; the people who were directly involved numbered in the thousands. In this article, I give an idiosyncratic account of the history, with a focus on the question of how people learned what measurement technology would be "sufficiently advanced" to succeed in detecting gravitational waves.

*The start of the search for gravitational waves*

Although the existence of gravitational waves was predicted by Einstein early in his development of the General Theory of Relativity,[2] Einstein himself never proposed that anyone search for gravitational waves. This contrasts with his well-known respect for experimental tests of his theory. He himself was immensely gratified that General Relativity explained the previously-measured precession of the perihelion of Mercury. The experimental confirmation in 1919 [3] of GR's prediction of the bending of starlight by the Sun made Einstein an international celebrity. The third of his three "classic tests", the gravitational redshift, was only successfully carried out in 1960. [4]

The reason for Einstein's silence on gravitational wave detection isn't known, but that is because there are too many reasons for it! No physical system, whether pre-existing in Nature nor purpose-built in the laboratory, could make a signal strong enough to be detected by any measuring apparatus conceivable in the early 20$^{th}$ century. (In his 1916 paper, Einstein remarks that the weakness of gravitational effects makes it difficult to conceive of a detectable gravitational wave.) Simply put, there was no prospect of any sufficiently advanced technology for the purpose.

But as if that weren't enough, Einstein also had serious doubts about the validity of his own prediction that gravitational waves existed. The mathematics of gauge freedom in General Relativity was slippery enough that the theory's inventor didn't see how to be certain that there was some gauge-invariant reality to the waves.

For these good reasons, there was no effort to search for gravitational waves for the first four decades after their prediction. When gravitational waves were discussed at all, it was in the context of trying to ascertain, by theoretical means, whether General Relativity predicted them or not.

All of this changed during the course of a single talk at the first-ever international conference on General Relativity. The Conference on the Role of Gravitation in Physics, held at the University of North Carolina at Chapel Hill on 18 – 23 January 1957, brought together 44 of the world's leading physicists who were working on problems related to General Relativity. While many of them arrived unconvinced of the reality of gravitational waves, a new consensus was established by the talk given by Felix Pirani, "Measurement of Classical Gravitation Fields." In it, Pirani gave a simple argument that made clear that something measurable happened when a gravitational wave passed through a set of free masses – there was an easy-to-calculate unambiguous change in the separations between the masses. Thus, at a stroke Pirani resolved Einstein's (and his successors') worries about whether there was anything coordinate-independent about gravitational waves; there was. (The history of this whole controversy is beautifully covered in Dan Kennefick's book, *Traveling at the Speed of Thought*.[5] For more on Pirani's central role, see [6].)

Pirani's talk was in the great Einsteinian tradition of resolving a conceptual question by proposing a thought experiment. What is remarkable about this thought experiment is that it led immediately to activity to realize it as an actual experiment. Joseph Weber was at the Chapel Hill Conference, in the company of his mentor John Archibald Wheeler. Weber must have been struck particularly by the exchange between Pirani and Hermann Bondi, during which Pirani said that the separation change between two test masses could be measured by sensing the stretching of a spring that connected them. By the time the Conference had ended, Weber had told the Conference's sponsor (Josh Goldberg, in his role with the U.S. Air Force) that he intended to follow up on Pirani's breakthrough. [7] Sure enough, within a few months a paper appeared by Weber and Wheeler, [8] recapping Pirani's argument (with suitable attribution), and explaining that one could in fact build a gravitational wave detector that implemented his idea. Over several years, Weber submitted essays to the Gravity Research Foundation essay competition, winning third prize in 1958, and first prize in 1959. The latter essay was also published in the *Physical Review*, [9] and constituted the manifesto of the research program that Weber pursued from that point forward, establishing the field of gravitational wave detection. Within about a decade, Weber was reporting that his detectors were finding evidence that gravitational waves were indeed being measured. [10]

There are many remarkable things about this story: the direct line (of almost zero length!) between the description of a thought experiment and the start of work to implement it, the (as it turned out) mistaken claims to have found signals, the world-wide effort to check Weber's discovery leading to its debunking, and the persistence (over decades!) of efforts to accomplish Weber's goal that finally led in 2015 to the genuine detection of gravitational waves and their establishment as a new means of doing gravitational physics and astronomy.

For the purposes of the present review, though, I'd like to highlight another seldom-remarked-upon aspect of the story. That is the fact that, when Weber embarked on his quest, he never seems to have asked the question about what sensitivity would be required to find naturally-occurring gravitational wave signals. He made the immediate leap from Pirani's demonstration of in-principle possibility to the campaign to realize that possibility. Decades before the film *Field of Dreams*, Weber bet his career on the maxim, "If you build it, they will come." [11]

Looking back over what it actually took to develop the "sufficiently advanced technology" (a team of scientists and engineers counted in the thousands, working for decades, using funds measured in units of hundreds of millions of dollars or Euros), Weber's confidence seems incredible. But before we call it madness, consider that his first detectors cost only a few tens of thousands of dollars, and could be built by a single principal investigator working with the help of only a few graduate students. Recall also that Weber was working when astronomy was being blown wide open by the creation of new observing modalities: radio astronomy was the most vivid example that would have inspired Weber in 1957, but through the course of Weber's development work in the 1960's, X-ray astronomy and infrared astronomy achieved major (and surprising!) discoveries. Why shouldn't he have been optimistic?

But while Weber was able to ignore the question at the start of his efforts, the issue was impossible to ignore once he started to report evidence for his reception of frequent gravitational wave burst signals. The situation can be understood by examining the 1972 review article "Gravitational-wave astronomy" by Press and Thorne. [12] The Introduction section contains the remarkable admission that, "We (the authors) find Weber's experimental evidence for gravitational waves fairly convincing. But we also recognize that there are as yet no plausible theoretical explanations of the waves' source and observed strength. Thus, we feel that we must protect this review against being made irrelevant by a possible 'disproof' of Weber's results." (I'd imagine that the authors soon wished that they could retract the "scare quotes" around the word "disproof.")

When discussing Weber's experiment specifically, Press and Thorne report that although it seemed just barely possible that a single one of Weber's bursts might have come "from a 'strong' supernova or stellar collapse somewhere in our Galaxy," they go on to note that "the number of bursts observed is at least 1000 times greater than current astrophysical ideas predict!" The next sentence, enumerating a rather desperate list of sources that might explain the apparent signals, ends with three question marks, followed by the statement that, "In fact, one is offered the tantalizing possibility that these new phenomena might dominate all other forms of energy generation and might force a major restructuring of our understanding of galactic and cosmological evolution."

Fortunately for physics and for standards of punctuation, the situation was soon resolved. Between 1972 and 1974, results came in from a number of experiments that had been undertaken to check Weber's results. [13] None of them found anything like Weber's results. (There was a single never-explained burst seen in coincidence between two bars operated by the Glasgow group. [14]) The Weber-induced frenzy was resolved by a conclusion (never shared by Weber!) that Weber's "events" were not genuine gravitational waves.

But the solution to one problem immediately led to a new problem: Should the search for gravitational waves continue, and if so, what level of sensitivity would be "sufficiently advanced" to allow success? Alongside these questions was another one: Would detectors of Weber's type (resonant cylinders with frequencies near 1 kHz and lengths of order 1 meter) constitute sufficiently advanced technology, or was altogether new technology required?

As with any difficult question, here different answers were given by different players. People moved forward from the Weber debacle in two main directions. A large portion of the community remained committed to Weber-style resonant-mass detectors, but aimed at a dramatic improvement in sensitivity through the use of cryogenic techniques. Another portion of the community (including many newcomers) embarked in a dramatically different direction – they aimed at much larger detectors (eventually several km in scale), with (effectively) free masses whose relative motions were sensed by tricked-up Michelson interferometers.

The subsequent decades of history tell us which idea was the one that eventually succeeded. But it is important to realize that the correct choice was far from clear in 1974. I want to describe each choice based on what was known, or could have been known, at the time. The question was, of course, what would constitute "sufficiently advanced technology" to enable the detection of gravitational wave signals. (A stimulating examination of this history can be found in *Gravity's Shadow* by Harry Collins.[15])

*The case for cryogenic bars*

It may seem strange to talk about technological conservatism in a field as ambitious as gravitational wave detection; and yet it made some real sense for people to stick with a technology that they understood. Of course, there had to be reason for optimism that sensitivity would improve substantially in a next generation instrument; after all, the community had just failed to see anything in their attempt to confirm Weber's claim.

Cryogenics held out the promise of the required big jump in sensitivity. Several factors combined in a virtuous way to give reasons for optimism. Recall that a resonant bar responded to a brief impulse by being set into vibration at its fundamental normal mode near 1 kHz. In order to be detected, that vibration has to stand out against the noise in the amplifier that senses the electrical signal generated by the transducer used to sense the vibration. It also needs to be stronger than the natural background of vibration of the mode due to Brownian motion. At low temperatures, very low noise amplifiers based on Superconducting Quantum Interference Devices (or SQUIDs) provide a much lower readout noise for small signals to compete against. Materials for the bar itself can have much higher quality factors (Aluminum 5056 alloy has a Q approaching $10^8$ at 50 mK),[16] allowing longer time averaging to further reduce the influence of amplifier noise. Finally, the lower temperature gives a direct reduction in Brownian motion noise, which is proportional to $k_B T$. One could expect that the noise could be brought down to an rms level equivalent to a strain of $h \sim 10^{-18}$, a full two orders of magnitude improvement in amplitude sensitivity beyond the best room temperature technology.

But would that be enough to meet the standard of "sufficiently advanced technology"? That depended on what signals were provided by the Universe. With a brand new observing modality, one could always be hopeful that there were heretofore unsuspected phenomena just faint enough not to have been discovered yet. The example of Galileo's new astronomy was often used -- Galileo was the first to turn a telescope toward the sky, and he was immediately rewarded with the discovery of sunspots, the moons of Jupiter, the phases of Venus, and the faint stars that make up the Milky Way. Could it not be the case that gravitational waves were like that? Weber had certainly seemed to think so. And perhaps he'd been just a bit unlucky in the level of sensitivity that he had reached.

On the other hand, it would have been much more of a sure bet if the new generation of detectors were aimed at a source that was known to send to the detectors frequent signals of sufficient strength to be found. In the 1970's, this kind of hope focused on gravitational wave bursts generated by gravitational collapse of the core of a massive star, the event that initiates one of the main kinds of supernova. The supernova rate was relatively well known, of order 1 (perhaps a few) per 100 years per galaxy like the Milky Way.

It was the strength of signals from supernovae that was hard to predict. At the upper end of optimistic predictions in the late 1970's, it was thought that perhaps they might generate a brief signal with an amplitude of about $10^{-18}$ at frequencies near the ~ 1 kHz resonant frequencies of bars, if the supernova occurred within our own Milky Way Galaxy. [17] So, optimism for this generation of detectors rested on a) achieving the new sensitivity target for cryogenic technology, b) supernovae being as "luminous" in gravitational waves as seemed at all possible, and c) an observing program that might need to last decades in order to catch one "golden" detectable event. This was a long shot if ever there was one, but at least it was something.

*The case for interferometers*

The idea of searching for gravitational waves with interferometers occurred to several groups of people independently. The first published proposal to use interferometers came from two Russian physicists, Gertsenshtein and Pustovoit.[18] The first interferometer to be built (on table-top scale) was by Weber's former student Robert Forward,[19] who had discussed the idea with Weber while they were still working together. But the most careful study of how interferometers might revolutionize the search for gravitational waves came from Rai Weiss, who in 1972 published (albeit only in an internal progress report of MIT's Research Laboratory of Electronics) a remarkable prospectus for the field.[20]

Weiss brought many key qualities to this task. A former grad student of Jerrold Zacharias and a former postdoc of Robert Dicke, he was looking for ways to apply precision measurement technology to problems in gravitational physics; he had also worked on developing early laser technology with Shaoul Ezekiel. Like many others, Weiss was intrigued by Weber's claims to have detected gravitational waves. Unlike others, he was interested in immediately leapfrogging to dramatically improved sensitivity, and thought that large (km-scale) interferometers were the appropriate technology for that purpose.

Weiss's reading of Weber's papers led him to the published papers of Felix Pirani. In the article in *Acta Physica Polonica*, [21] unlike in Pirani's talk at Chapel Hill, Weiss read that Pirani described his thought experiment not as masses that might be connected by a spring, but instead as spatially separated masses whose separations could be monitored by measuring the round-trip time of light signals that traveled between them. Weiss recognized that a Michelson interferometer performed just such a travel time measurement, actually a comparison of two round-trip travel times on perpendicular paths. This was a) perfectly suited for the quadrupolar strain pattern of gravitational waves and b) embodied the principles of differential measurements of which Dicke was so fond.[22] (In this case, the balanced instrument geometry nulled out what would otherwise be a large noise term, fluctuations in the laser light's frequency.)

Of course, unlike Michelson's interferometer, one that was aimed at detecting gravitational waves would need the mirrors to be more-or-less free, not bolted down. This called for an elaborate feedback control system, more grist for the mill of techniques from the school of Dicke.

By 1972, with Weber's claims attracting attention everywhere, Weiss wrote his prospectus for a future program of interferometric gravitational wave detectors. In it, he was agnostic on the validity of Weber's claims. Intriguingly, Weiss took as his target source the steady nearly-sinusoidal gravitational waves emitted by the Crab Pulsar (if it is sufficiently far from axisymmetric.) Note, though, that since the Crab Pulsar's eccentricity wasn't known (and still isn't, beyond some ever-improving upper limits [23]), Weiss could not use it for the purpose of establishing the sensitivity that would be required of an interferometric gravitational wave detector.

Weiss pointed out that size really matters in gravitational wave detectors. Because the wave's effect on a set of test particles is to apply a (quadrupolar) strain, the actually separation change between any pair of test masses is proportional to their separation. Resonant detectors have their lengths determined by the need for their fundamental resonant frequency (the inverse of the round-trip sound travel time between the ends of the detector) to lie in a band containing substantial power from the wave itself; given the speed of sound in materials and the estimated signal frequency of about 1 kHz, that led directly to the size of order one meter that was characteristic of Weber's detectors and all that were developed subsequently. Interferometers had the promise to benefit from much larger separations, until the round trip light travel time becomes comparable to the signal period. The fact that the speed of light is so much greater than the speed of sound in materials thus shows that interferometers have a natural advantage of many orders of magnitude.

Perhaps the most remarkable feature of Weiss's 1972 paper is the extensive and detailed analysis of the noise budget of an interferometer. Workers in the field today are still astounded at how large a fraction of the essential difficulties were understood already at the start by Weiss. This analysis confirmed "from the bottom up" Weiss's top-down argument that large interferometers would have an advantage over Weber-style detectors in direct proportion to the ratio of lengths. Almost every term in the noise budget (as given in units of displacement) had a strength that was independent of length, so that the signal to noise ratio (to a signal whose strength was characterized by a dimensionless strain) improved with length. (The only exception was Newtonian gravitational noise from density fluctuations in the environment, and then only in the low frequency limit where seismic/acoustic wavelengths are long compared to the size of the detector. But this noise source is only strong at frequencies around or below 10 Hz, and fell steeply at higher frequencies.)

We can mention a frequency of 10 Hz in this context because, unlike Weber-style resonant detectors, interferometric detectors are in principle (and in fact) capable of broad-band operation, where the output of the instrument gives a faithful readout of the strain waveform $h(t)$ of the incident gravitational wave. Weber-style detectors use the high-Q resonance to boost signal strength above the sensor noise in the displacement readout, by averaging for a long time. But, to the extent that the test masses in an interferometer are more-or-less free, the bet could be made that displacement readout noise (in this case, shot noise in the laser light) is small enough that one can sample the strain output many thousands of times per second.

The actual useful bandwidth of an interferometer is determined by a few factors.

- The "free" test masses aren't truly free, but are masses suspended as pendulums with a resonant frequency of about 1 Hz. Thus, performance will be limited for frequencies near 1 Hz and below.
- Several noise terms (including Newtonian gravitational noise, the (usually) much stronger direct seismic noise coupling, as well as other technical effects) tend to be much stronger at low frequencies than at high frequencies. Thus, the noise budget tends to have a "wall" preventing operation at or below frequencies of order 10 Hz.
- There's a gradual roll-off of sensitivity at high frequencies, from the competition between frequency-independent shot noise and the frequency-dependent strain sensitivity.

Still, history has since shown that useful observations can be obtained over a broad band between ~ 10 Hz and a few kilohertz, with the possibility of searching for a wide range of gravitational wave signals and also with faithful reproduction of the signals' waveforms once they are found.

Compared with all of these "on paper" benefits, it was clear even in 1972 that there was a pair of daunting drawbacks. Everyone who read Weiss's prospectus could see that what he was proposing was both awesomely complex and much more expensive than Weber-style detectors. Both of these drawbacks proved to be correct descriptions of the detectors that succeeded decades later.

*Taking stock at the Battelle Conference, 1978*

Key members of the gravitational wave detection community gathered in late summer of 1978 at the Battelle Seattle Research Center to consider the state of the field, in a workshop called Sources of Gravitational Radiation.[24] Notable especially was the focus on trying to establish believable estimates of the strengths of signals that would be presented to the next generation of detectors. The workshop was bookended by two discussion sessions, the first on the current state of detection technologies and on their prospects, and the second devoted to comparing likely instrumental sensitivities with the strengths of predicted signals. The published proceedings of the workshop include transcripts of comments made in the discussion sessions, thus allowing deep insight into the state of the field. During the discussion session on detectors, Robert Forward is quoted as saying, "At least we are now able to draw the antenna sensitivity curves and the source [strength] curves on the same graph. Surely [laughter and applause] this means we have come a long way."[25]

That quote indicates that the field had matured quite a bit in the previous decade. Recall that the 1972 review article by Press and Thorne contained a wild mix of sober and highly speculative ideas for gravitational wave signal strengths, motivated by the then-still-live possibility that Weber's events were real. By 1978, participants knew that Weber's "events" could be disregarded, and wanted only sober assessments of what they needed to be looking for.

Just as Pirani's talk at the Chapel Hill Conference of 1957 resolved theoretical questions about the physical reality of gravitational waves, there was a talk at the Battelle Workshop of 1978 that resolved questions about what would be the most-likely-to-be-detected source of gravitational waves. But unlike the case Pirani's 1957 talk, which instantaneously convinced everyone who heard it that it was a "game changer", the talk given by J. Paul A. Clark only gradually came to be recognized as the essential contribution that it was.[26] Clark had been a graduate student of Douglas Eardley at Yale, where the two of them wrote one of the early papers describing the gravitational wave signal from the inspiral and coalescence of a neutron star binary. [27] That paper concluded that signals from binary coalescences would be "less important than supernovae as sources of gravitational waves." However, by the time of the Battelle Workshop, Clark had realized that that conclusion was both premature and unduly pessimistic. Firstly, it now looked as if the abundance of neutron star binaries was much higher than he had previously believed (by nearly two orders of magnitude.) Secondly, the previous comparison with supernovae had assumed that supernovae were efficient sources of gravitational waves, but that was coming to seem less certain. The resulting conclusion was that, unless supernovae were indeed as efficient as anyone had thought to be at all plausible, neutron star binaries would more important sources of gravitational waves. (Oh and, by the way, something similar was true for mixed neutron star/black hole binaries as well!)

Perhaps the impact of this paper was blunted by Clark's choice to present this news in terms of a comparison with the more-uncertain-by-the-day supernova signals. But gradually, readers came to understand that Clark's conclusion was even more important if expressed without any reference to supernova signals. What he had shown was that:

- neutron star binaries were efficient sources of gravitational waves whose waveforms (including their strengths) could be calculated directly from general relativity,
- the abundances of neutron star binaries were reasonably well established, leading to rather secure predictions of the amplitudes of signals that would arrive at gravitational wave detectors at any given detection rate (say, once per year),
- the waveforms of compact binary signals was such that it would be much more favorable to search for them with broadband detectors sensitive below 1 kHz than it would be to use resonant detectors sensitive just at 1 kHz, and

- the peak amplitude of the likely once/year signal was unfortunately somewhat below $h \sim 10^{-21}$, more probably $h \sim 10^{-22}$.

As this news was digested by the gravitational wave community, its impact became clear. If one wanted to be sure to have designed a "sufficiently advanced" detector, the target source was now known: neutron star binaries (or their close cousins involving black holes.)

The impact of Clark's talk (which, like Pirani's work decades earlier, had been submitted to a regular journal for publication prior to his delivery of the talk [28]) was different on the resonant detector community than it was on the nascent group of people developing interferometers. For the resonant detectors, Clark's signal wasn't all that well suited. Thus, their target source continued to be a supernova in our Galaxy, hoping that it was as strong as it could possibly be, and also hoping that one didn't have to wait 50 years for one to go off. For interferometers, the source seemed almost tailor-made. But on that word "almost", a tale hangs, as I'll describe below.

*The "Blue Book" design study of large interferometers*

For too many of us (at least in the American context), there's a reflex to think of government as at best a non-creative influence and at worst as something that actively interferes with creativity. Nothing could do more to counter such attitudes than to study the role of the U.S. National Science Foundation in fostering the discovery of gravitational waves.

If there is a single hero to this part of the story, it is the physicist Richard Isaacson (although he in turn is quick to credit his mentor at the NSF, Marcel Bardon.) Isaacson, a gifted relativist in his own right,[29] served as the NSF's Program Officer in Gravitational Physics from 1973 to 2002. His energetic and insightful leadership, out of public view but very much visible to members of the community that he nurtured and supported over three decades, is as responsible as any other individual's contribution for the success of gravitational wave detection.

When Isaacson first took on leadership of NSF's portfolio of gravitational wave research, the search for gravitational waves was at a fever pitch of excitement and turmoil. The attempted replications of Weber's claims was beginning to coalesce in the community's rejection of those claims. The question of what should come next was on everyone's mind, as we saw above. Isaacson was determined to do what he could to support that community and to ensure that it carried out the best science possible.

For the portion of the community focusing on resonant detectors, Isaacson offered substantial support for the development of the new generation of cryogenic detectors. He also insisted that the community bring the development work to fruition as soon as was practicable, and that it resist the seduction of "sweet" technology in favor of a focus on actually searching for gravitational waves. By the time of the 1978 Battelle workshop, several groups' detectors were almost ready to turn on. The start of the 1980's saw the publication of the first observations; for the rest of the '80's and '90's, Isaacson helped to ensure a healthy program of instrument development and observation. Quietly (and for the most part behind the scenes), he also insisted that the scientific standards of the groups' observations be the highest possible; blind analysis was required, and the continued evidence of the rejection of Weber's never-surrendered claims was highlighted at every opportunity.

For those who were just beginning the development of interferometric detectors, Isaacson necessarily offered a different kind of support. He saw clearly that interferometric technology had great promise but that it was far from ready to implement. In part, the lack of readiness was strictly technical: no one had operated an interferometer at anywhere close to the sensitivity that would be required. But just as important in Isaacson's mind was the lack of a community ready to design, engineer, build, commission, and operate a highly complex detector, far larger in scale and budget than anything that the gravitational wave community had ever done (or even than NSF had ever supported!)

Under the leadership of Kip Thorne, Caltech made a major commitment (supported by Isaacson at NSF) to develop interferometric gravitational wave detectors, and in 1979 succeeded in recruiting Ronald Drever from Glasgow to lead the experimental effort. The highlight of the investment there was the construction of a prototype interferometer with arms 40 meters in length.

Isaacson carried out a different sort of community building with Weiss at MIT; there, NSF funded a proposal to carry out a design study that examined whether it would indeed be possible to scale up an interferometer from the small prototypes then in existence to the multi-km scale at which the technology would fulfill its scientific promise. [Personal note: The grant included funds for a postdoc to support the design study, who turned out to be me, in my first job after earning my Ph.D.] Isaacson saw, before anyone else, that it would be just as important to present competent engineering as to present a strong science case or convincing laboratory-scale technology demonstrations, if the large interferometers were going to receive the substantial funding that would be required. Thus, NSF's grant to MIT explicitly called for working with industrial engineering firms to validate that, say, the huge vacuum system could be constructed successfully and operated for a reasonable amount of money.

No other form of gravitational wave detector showed any believable chance of reaching the level of sensitivity that would be required to detect the strongest guaranteed source, Clark's neutron star binaries. But could interferometers in real life fulfill their on-paper promise to meet of sensitivity to strains of $10^{-21}$ or smaller? Isaacson's challenge to the Caltech and MIT groups was nothing less than to demonstrate that "sufficiently advanced technology" could truly be developed for the task of detecting gravitational waves.

The MIT design study (which came to be known as the "Blue Book" because of the color of the covers used when it was presented to the NSF) was completed in October 1983 (with a contribution also from Stan Whitcomb at Caltech.)[30] It reported that the engineering requirements of large interferometers could indeed be met, for a then-considered-large cost of order $100 M (still substantially below what turned out to be the true cost.) The document reviewed the likely sources of signals, with content similar to that of the Battelle Workshop proceedings; the "guaranteed" signal comes from neutron star binaries. The crucial sentence is,

"To see around 10 events per year we must see to distance of 150 Mpc, where maximum amplitudes are around h=$10^{-22}$."

The section of the report describing expected instrumental performance was remarkably optimistic. Based on several aggressive assumptions about what might be feasible

- 100 W of laser power,
- active seismic isolation, and
- arm cavity storage time optimized at each frequency (which could have made sense if described as the result of the use of Fabry-Perot cavities in the arms, even though that was not specifically mentioned), the report predicted a noise spectrum remarkably close to that of the Advanced LIGO instruments on 14 September 2015. There was also another slightly less optimistic noise spectrum for arms with fixed storage time of 1 ms, corresponding to arms containing Herriott delay lines, the favorite design of the MIT authors.

The Blue Book was presented late in 1983, jointly by Weiss's MIT research group and Drever's Caltech research group, to the NSF's Advisory Committee for Physics. It was received respectfully, giving Isaacson the endorsement he needed to keep funding the development of interferometers.

*Brief apology to the reader*

For the history narrated to this point, I believe that I've covered the highlights of the field without regard to where the events took place. The subsequent history is more complicated, and took place in multiple parallel tracks (primarily) in the U.S. and in a number of European countries. The author knows the U.S. history fairly well, but is not capable of doing justice fully to the parallel developments in Europe. Thus, in the remainder of this review, events in the U.S. will be featured, with (unfortunately) only brief mentions of the similar progress taking place elsewhere. A future review by a more knowledgeable person will be needed to fill in the portions of the history that this review is unable to cover.

*LIGO's 1989 construction proposal*

Over the next half-decade, the Caltech and MIT groups struggled in their respective laboratories with prototype interferometers, and struggled with each other to invent a plan that could be formally proposed for construction. (Its name was also coined, as the Laser Interferometer Gravitational-Wave Observatory, or LIGO.) By 1989, the two groups were jointly led by Rochus "Robbie" Vogt, a gifted experimental physicist who had until recently been the Provost of Caltech. The Caltech group had also made good progress on its 40 meter long prototype interferometer, achieving encouraging (if not yet quite sufficient) displacement sensitivity at 1 kHz.

Under Vogt's leadership, on 1 December 1989 the combined groups submitted to the NSF a proposal for a pair of 4 km long interferometers at sites widely separated within the United States.[31] By any measure, the plan was audacious; the plan called for funds larger than for any project ever yet supported by the National Science Foundation. It also assumed that it would be possible to scale up the sensitivity far beyond any gravitational wave detector ever yet constructed, whether Weber-style or interferometer. Vogt was moved to start the proposal with an epigraph quoting Machiavelli's *The Prince*, "There is nothing more difficult to take in hand, more perilous to conduct, or more uncertain in its success, than to take the lead in the introduction of a new order of things."

And yet, for all of its aggressive optimism, the interferometers that were proposed to be constructed were not "sufficiently advanced technology" to have any assurance of success. In fact, the design sensitivity would fall about one order of magnitude short (in units of strain amplitude or of distance to be searched) of what would be required to be assured of finding neutron star binary signals.

What had happened? The team had convinced itself that it was going to be hard enough to achieve even these scaled back goals, and that they didn't have a plan to achieve the more aggressive design that had been imagined in 1983. A couple of examples:

- Laser power was projected to be 5 W (not 100 W), still a large scale-up from what had been used in the laboratory prototype interferometers.
- Seismic isolation goals were also scaled way back; LIGO would use a simple multi-stage passive vibration isolation system, not the active system envisaged in the Blue Book.

Even thus scaled back, the project would not only be an enormous undertaking, it would represent a truly dramatic improvement in gravitational wave sensitivity – about three orders of magnitude (in strain units) beyond the best previous observations (those of the most recent generation of cryogenic resonant detectors.)

Still, given what was known about sources of gravitational waves, how could this scale-back in design sensitivity be justified? The answer was twofold: a) by appealing to what <u>wasn't</u> known about sources of gravitational waves, and b) by making clear that the proposed instruments ought to be succeeded by more advanced instruments that would (it was hoped) make up the gap in sensitivity between the first generation's "best feasible" sensitivity and what would be "sufficiently advanced" to be guaranteed to succeed.

The clearest explanation of this strategy was given on pages 8 – 10 of the 1989 construction proposal, in a section called "Estimates of the Strengths of the Waves at Earth and Comparison With Anticipated LIGO Sensitivities." Figure II-2 of the proposal (shown in the figure below) is an information-rich combination of projected instrument sensitivities versus frequency, combined with projected strengths of various sources. (This is a descendent of the graphs that the participants at the Battelle Workshop had been so proud to have constructed for the first time.) On it are shown three levels of instrument performance: a) a curve labelled "Prototype" showing the best performance of the Caltech 40-m interferometer, b) a second one labelled "LIGO Early Detector", showing the expected performance of the instrument proposed for construction in the current proposal, and c) a curve labelled "LIGO Advanced Detector" showing something like the Blue Book's most optimistic curve, substantially better at all frequencies than the detector that was about to be built. (It was close to an order of magnitude more sensitive from 100 Hz upwards, and had good sensitivity down to frequencies about an order of magnitude lower.)

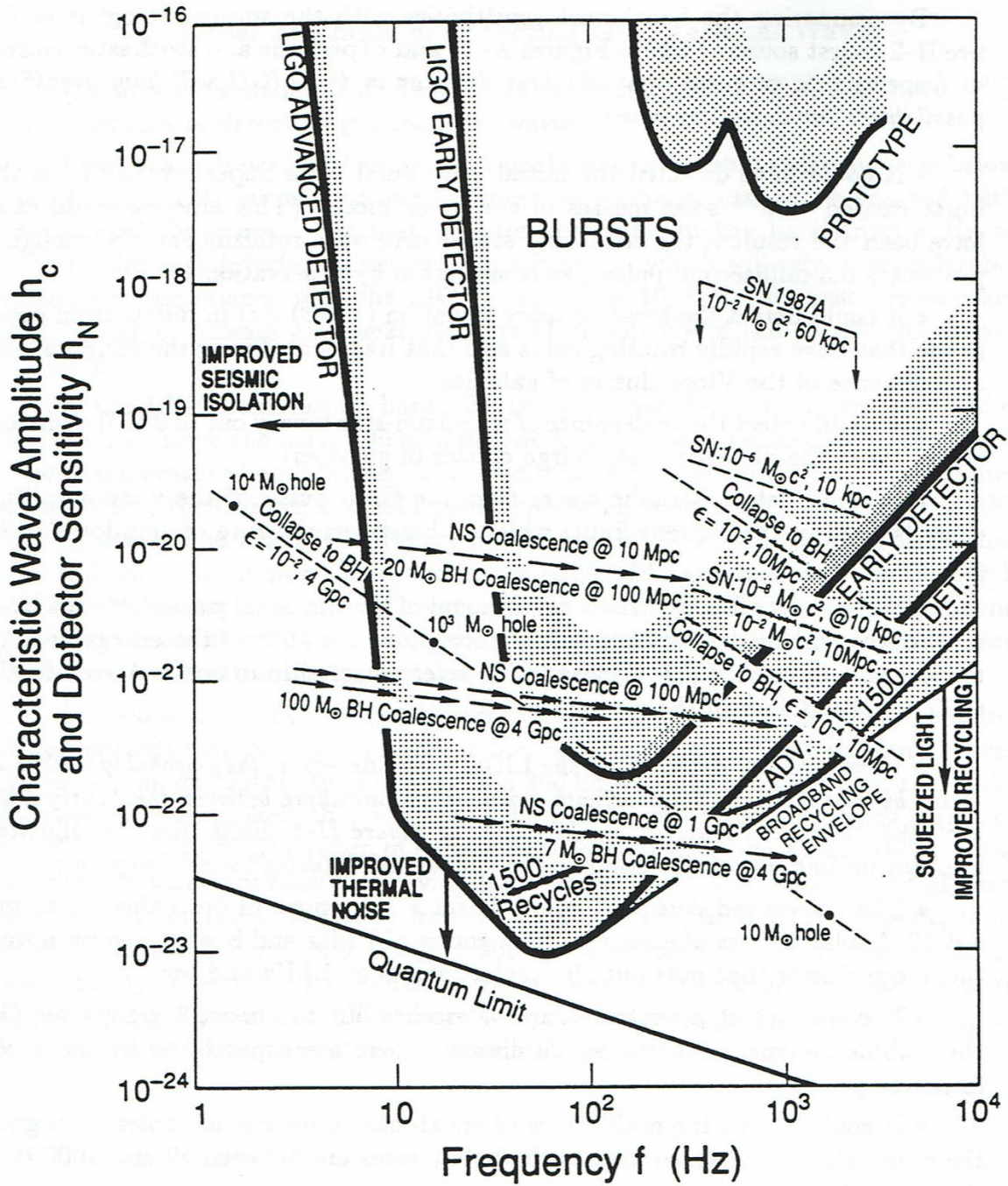

Figure 1: A reproduction of Figure II-2 from the 1989 LIGO construction proposal.

In the text accompanying this figure, it is explained that the Early Detector is what is being proposed in this proposal. It goes on to say, "Once the early detector has been operated near the sensitivity of the middle curve, *a succession of ever-improving detectors will evolve*, [emphasis mine] continually pushing the sensitivity level down (to smaller h) and left (to lower frequencies, f). As a rough indication of where this might lead after a few years, we show … the sensitivity of an 'advanced detector'".

Completing the discussion of the science case, the section then comments on the relationship between these sensitivity curves and the strengths of signals from different sources. Using words carefully, it states that "*the first detector in the LIGO will have significant possibilities for detecting waves* [emphasis in original]." It goes on to mention four possibly detectable signals, among them:

- "Supernova 1987A if that burst carried ~$10^{-4}$ solar masses of energy or more" (today considered an optimistic efficiency,) and
- "the coalescence of a neutron-star binary out to 30 Mpc distance"

This latter was a factor of 5 or so below what was thought needed to guarantee success, but note that GW170817 was detected on 17 August 2017 at a distance of 40 Mpc.

The section goes on to say that "If waves are not detected by the LIGO's first detector, *they will probably be detected by a subsequent detector with sensitivity somewhere between the "early detector" and "advanced detectors" sensitivities of Figure II-2.*" [emphasis in original] There follows a list of five possible signals, among them:

- "a supernova in the Virgo cluster that puts out $10^{-2}$ solar masses at 1 kHz and less" (choosing a distance that would include frequent enough signals, but still assuming much too high a gravitational wave emission efficiency according to today's best estimates),
- "a neutron-star coalescence out to almost 1 gigaparsec" (this is, finally, beyond what would be needed), and
- "It could detect the coalescence of equal-mass binary black holes throughout the observable universe, so long as the hole masses are between 10 and 1000 times the mass of the Sun." (With the benefit of hindsight, this was the most exciting prediction of all.)

Looking back now, it is truly remarkable that a proposal so honest about its slender prospects was approved! Evidently, the excitement of a potential gravitational wave discovery was strong enough to overcome the meager chances of near-term success. One can only applaud the courage of everyone involved (the proposers, the reviewers, and the funders.)

But we need to be clear that this meant a very long timescale before success could be expected. The LIGO proposal was approved by the NSF in 1990, and received its first funds from the U.S. Congress in 1991. It wasn't until 2002 that the first preliminary observation run took place, with design sensitivity (of the "initial" version of LIGO) being reached in 2005. Only after the five year period of design sensitivity (and better) observations were carried out could the observatories be made available for the successor instrument, which by then had come to be called Advanced LIGO.

In spite of this author's focus on developments in the United States, it is important not to lose sight of the fact that the field was developing in parallel in Europe as well. In 1989, two other interferometer construction proposals were submitted: the German-British GEO proposal and the French-Italian Virgo proposal. Each was to have 3 km arms. Many details differed, with much of today's best technology (for example, in lasers and in vibration isolation) prefigured in these proposals. The two proposals in Europe met different fates:

- In spite of its merits, the GEO proposal couldn't be funded once Germany re-unified, calling for a re-orientation of national budgets toward national reconstruction and integration. Instead, the GEO team found funds to build a 600 meter interferometer. Besides conducting observations, GEO600 became the premier site for development of several advanced technologies, for example squeezed light.
- The Virgo proposal was funded. Virgo progressed on close to the same timescale as LIGO, observed alongside LIGO during the "initial detector era", and participated (among other things) in the discovery of the neutron star binary signal GW170817.

Each of the two European teams also forged excellent working relationships with LIGO, each in its own way.

*The road to the advanced detectors*

In 1994, leadership of the LIGO Project passed from Robbie Vogt to Barry Barish, a very experienced leader in high energy physics. Barish led major re-organizations of all aspects of the LIGO Project. By 1997, he carried out his most far-reaching re-organization with the creation of the LIGO Scientific Collaboration, a body of university-based scientists around the U.S. (and eventually many other countries as well), who were to "carry out the scientific program of LIGO."[32] The LSC included what had now been constituted as the LIGO Laboratory (the group of scientists and engineers at Caltech, MIT, and the two LIGO sites who were responsible for the interferometers themselves,) but was open to all other scientists interested in making a commitment to LIGO's science and technology.

The NSF, under Isaacson's continued guidance, had already been funding a number of independent research groups who were doing LIGO-related research without any formal connection to the LIGO Project itself. Barish and his team initially coordinated with that group of scientists via informal channels, leading up to a short-lived coordinating body called the LIGO Research Community. But NSF needed to better demonstrate how the scientific benefits of LIGO would accrue to a broad community beyond the two elite institutions who were managing the project. In July 1996, Isaacson brought together a blue-ribbon panel to consider the Use of the Laser Interferometer Gravitational-Wave Observatory. Barish consulted with panel chair Boyce McDaniel (a longtime colleague from high energy physics), and the two delivered the solution that has served the community very well up until the present moment.[33] Rai Weiss was appointed to serve as the first Spokesperson of the LSC.

The work of the LIGO Scientific Collaboration focused on two main areas:

- carrying out the analysis of data in the search for signals, and
- developing technology for detectors more advanced than the initial LIGO interferometers.

With no disrespect intended (at all!) to the data analysis work, here I'll focus on the work directed toward advanced detector technology. It was this effort that enabled the LIGO team (Lab plus LSC) to make good on the promise that the LIGO instruments would eventually be able to operate at sensitivity levels that would guarantee a detection.

The LSC held its first meeting in August 1997 at Louisiana State University. Among its first tasks was to see if it was possible to create a credible design for an interferometer that would meet the 1989 proposal's standards for an "advanced" detector. Fortunately, in the intervening years there had been a great deal of technological progress. Within two years, the LSC had produced a "White Paper on Detector Research and Development",[34] containing a "reference design" that would, on paper at least, meet the need.

It was not universally agreed whether the advanced detector would gradually be approached through a series of partial upgrades of individual subsystems (as suggested in the 1989 construction proposal) or whether all improvements should be brought together at once in a complete replacement of the original interferometers. The broader physics community had a major influence on this decision, through the deliberations of the Committee on Gravitational Physics. This body, chaired by Jim Hartle, was established to contribute a gravitational physics section to the Decadal Survey for Physics being assembled toward the end of the 1990's by the National Academy of Sciences. The CGP devoted a not inconsiderable amount of its attention to LIGO. The report [35] was dramatically supportive of LIGO. In its list of "opportunities for the next decade", the first three items were things that would come from successful LIGO observations:

- "The first direct detection of gravitational waves by the worldwide network of gravitational wave detectors now under construction.
- The first direct observation of black holes by the characteristic gravitational radiation that they emit in the last stages of their formation.
- The use of gravitational waves to probe the universe of complex astronomical phenomena by the decoding of the details of the gravitational wave signals from particular sources."

In the list of "goals", the first two items listed were also focused on gravitational wave detection:

- "Receive gravitational waves and use them to study regions of strong gravity."
- "Explore the extreme conditions near the surface of black holes."

Finally, under "Recommendations", the first to be listed were Gravitational Waves, where, under the subheading of "The High-Frequency Gravitational Wave Window", were these three specific recommendations:

- "Carry out the first phase of LIGO scientific operations.

- Enhance the capability of LIGO beyond the first phase of operations, with the goal of detecting the coalescence of neutron star binaries.
- Support technology development that will provide the foundation for future improvements in LIGO's sensitivity."

And in the text supporting those recommendations, the report states that the CGP "recommends support for *sustained* [emphasis in original] development of the technology necessary to upgrade LIGO to a sensitivity necessary to detect neutron star binary coalescences."

No one could have asked for stronger support from the community! There was also extremely helpful guidance from the CGP that was conveyed informally to me during a dinner meeting of the committee. (I was privileged to serve as a member of the CGP.) Clearly, LIGO's success was a matter of great concern to the Committee's members. But so too was LIGO's large budget. Members made it clear that the community could only support one large upgrade, and it would have to succeed as soon as possible. No one on that panel liked the idea of a long program of incremental improvements. They urged strongly that the LIGO team ought to make one complete well-planned upgrade that would enable the achievement of "sufficiently advanced technology" as soon as could be done.

By the time that the 1999 White Paper on Detector Research and Development was written, the private guidance from the CGP had been adopted as LIGO's plan. (Fortunately, that guidance fit well with the opinions of many technical experts within LIGO about how much easier it would be to implement a broad set of changes all at once, as opposed to sequentially.) The White Paper presented a vision (called a "reference design" for what was then called "LIGO II") for a single replacement for the initial LIGO instruments, not a series of adiabatic improvements aimed to reach a new goal asymptotically. The proposed improvement is ambitious, roughly a factor of twenty more sensitive in range.

Several things stand out about this new vision. Firstly, even though it was ambitious, it was realistic (at least by the standards of this ambitious field of science.) The decade that had elapsed since the 1989 construction proposal had indeed seen great progress. A large step upward in laser power (180 W in the reference design, compared with 10 W for initial LIGO) seemed within reach. A dramatic improvement in pendulum thermal noise was envisioned, from the use of fused silica instead of steel for the pendulum fibers; a technical breakthrough in realizing this long-desired design had come from glass-joining technology developed for the Gravity Probe B satellite.[36] Test mass thermal noise was also slated for a big reduction; sapphire test masses were suggested, but there was explicit mention of fused silica as a fallback, with knowledge by then of how well fused silica could perform if the correct design and construction choices were made. And there were plans for much better seismic isolation for reduction of noise at the lowest frequencies, with mention of two competing visions for how that could be achieved.

Equally striking, and more relevant to the point of this essay, was how explicit the White Paper was in its emphasis on signals from compact binary system as the justification for the design. In the section called "Physics Reach", just after the "Summary and Introduction", the first five possible signals mentioned are all variations on compact binary coalescences. The performance projected is indeed impressive – ranges out to redshifts $z$ of a few tenths! The authors refrained from stating explicitly that this is what would make all the difference between the "crap shoot" that was initial LIGO and the "sure thing" that was being promised here; instead of a statement of how large a range was required, there was a reference to an essay by Thorne reviewing the expected abundances of sources. Likely this was out of some residual hope that initial LIGO would come through after all – this was 1999, still years before the first LIGO science run and even more before initial LIGO reached its design sensitivity. But for anyone informed of the situation, including LIGO's anxious friends in the larger gravitational physics community, the envisioned LIGO II would be a godsend.

(Notable also is the complete absence of the word "supernova" from the list of possible sources. This is a bit misleading, though, since the source mentioned right after the list of five versions of compact binary signals is "r-mode oscillations of a newborn neutron star", at the time the most respectable idea for a strong gravitational wave signal from supernova core collapse.)

Most remarkably of all, this White Paper redeemed the nearly-incredible strategy of the 1989 LIGO construction proposal (and of its siblings in Europe), which justified a scientifically inadequate design by giving a promise that we'd figure out how to do better in time to propose an adequate successor instrument later on.

The 1999 LSC White Paper served as the prospectus for further research in the U.S., allowing the LIGO Laboratory to submit a proposal in 2003 for the construction of Advanced LIGO. Note that this was still two years before initial LIGO had achieved its full design sensitivity! But given the community's strong support as well as the long lead time it takes for projects of this scale, the aggressive timing of this proposal was sensible. (A revised proposal was submitted in 2005; it was approved by the National Science Board in 2008, after initial LIGO had achieved its design sensitivity and had carried out a long observing run.)

*Conclusion*

Looking back from the current vantage point, it looks even wiser to have pushed so early and so aggressively for what became Advanced LIGO. It took until late 2010 for the start of installation of Advanced LIGO in place of the initial LIGO interferometers, and until 14 September 2015 for the first detection of a binary black hole coalescence signal during aLIGO's first observing run. After decades of "irrational exuberance",[37] during which the search for gravitational wave signals was conducted with detectors whose sensitivity was far below what was actually believed to be required, success came when instruments known to be (close to) adequate finally came on line. And yet, it is surely not my intention to criticize the field for having persevered without any realistic near-term hope of success. After all, it is hard to conceive of any path that would have led to the current success of the LIGO and Virgo interferometers other than the one that was followed. We should have feelings of gratitude for the steadfastness of those in the field and those who funded the work throughout the decades that were required to develop sufficiently advanced technology.